\documentclass[showpacs,showkeys,preprint,amsmath,amssymb,superscriptaddress]{revtex4}    
\usepackage{graphicx}
\usepackage{dcolumn}
\usepackage{bm}
\newcommand{\be}{\begin{equation}}
\newcommand{\ee}{\end{equation}}
\newcommand{\bee}{\begin{eqnarray}}
\newcommand{\eee}{\end{eqnarray}}
\newcommand{\eq}{\end{quote}}
\newcommand{\nn}{\nonumber}
\newcommand{\Slash}[1]{\ooalign{\hfil/\hfil\crcr$#1$}}

\def\lsim{\displaystyle\mathop{<}_{\sim}}

\begin{document}      

\preprint{PNU-NTG-06/2003}
\title{Photoproduction of $\Theta$ baryon from the neutron}
\author{S.I. Nam}
\email{sinam@rcnp.osaka-u.ac.jp}
\affiliation{Research Center for Nuclear Physics (RCNP), Ibaraki, Osaka
567-0047, Japan}
\affiliation{Department of
Physics, Pusan National University, Pusan 609-735, Korea}

\author{A. Hosaka}
\email{hosaka@rcnp.osaka-u.ac.jp}
\affiliation{Research Center for Nuclear Physics (RCNP), Ibaraki, Osaka
567-0047, Japan}

\author{H.-Ch. Kim}
\email{hchkim@pusan.ac.kr}
\affiliation{Department of
Physics, Pusan National University, Pusan 609-735, Korea}
\date{\today}

\begin{abstract}
We study photoproduction reactions for $\Theta^+$, which was
recently observed as a candidate of an exotic five quark state
$uudd\bar{s}$. We employ effective lagrangian interactions in the
linear and non-linear formalism of chiral symmetry and investigate two
cases of spin and parity of $\Theta^+$, 
$J^P = 1/2^{+}$ and $1/2^{-}$. We
calculate cross sections and discuss possibilities to discriminate one
from the other. 
\end{abstract}

\maketitle

\section{introduction}
Recent discovery of the pentaquark baryon $\Theta^+$ with strangeness
$S=+1$ has triggered a diverse interest in a new type of hadron
containing more than three quarks. 
Experimental evidence was first reported by
LEPS group at SPring-8~\cite{Nakano:2003qx} in a reaction $\gamma n \rightarrow
K^{+}K^{-}n$. This result was subsequently confirmed by 
DIANA (ITEP)~\cite{Barmin:2003es}, CLAS (J-Lab) and
SAPHIR~\cite{Barth:2003es} collaborations.  
So far the mass and width of $\Theta^+$ were
observed with a relatively small mass $M_{\Theta}\sim 1540$ MeV and a
very narrow width $\Gamma \lsim 25$ MeV. 
Experimentally, other properties are not yet known. 
On the contrary,
there are several theoretical studies for some time using various
effective models~\cite{Walliser:vx,Diakonov:1997mm,Lipkin:1998pb}.  
Also, there are numbers of preprints that attempt to explain  the nature of 
$\Theta^+$~\cite{Stancu:2003if,Capstick:2003iq,Hosaka:2003jv,Jaffe:2003sg,Wybourne:2003if,Karliner:2003sy,Walliser:2003dy,Kelkar:pt,Zhu:2003ba,Nussinov:2003ex,Borisyuk:2003fk,Glozman:2003sy}. 
Perhaps the most attractive predictions were made by
Diakonov et al. in the chiral soliton
model~\cite{Diakonov:1997mm}. Using a few input 
parameters, the masses and decay widths of various channels were
calculated with remarkable agreement with experimental data. In this
model, $\Theta^+$ was identified with a member of flavor antidecuplet
$\bar{10}$ with $S=+1$, isospin $I=0$ and spin-parity
$J^{P}=1/2^{+}$. 
A peculiarity of these predictions is the positive parity of $\Theta^+$,
since in a naive valence quark model, the ground state of the five
quark state must have negative parity. 
There are several suggestions which brings down the $1/2^+$ state 
lower than the $1/2^-$
state~\cite{Stancu:2003if,Hosaka:2003jv,Jaffe:2003sg,Glozman:2003sy}.   
For instance, in Ref.~\cite{Hosaka:2003jv}, the role of the hedgehog pion  
was emphasized, while in Ref.~\cite{Jaffe:2003sg}, diquark correlations
were investigated.  
Definitely, more theoretical and 
experimental studies are needed in order to
further understand the nature of $\Theta^+$. 

The purpose of this paper
is to provide a theoretical estimate for photo-induced 
$\Theta^+$ production reaction
$
\gamma n \to K^- \Theta^+, 
$
and present various cross sections which will be utilized in further
experiments. 
In actual experiments, 
the final state $\Theta^+$ decays into $KN$ eventually.  
Here, we do not consider this part for the following reasons.  
First, the fact that the observed peak of the resonance is 
very sharp with a narrow width implies that at least in the 
vicinity of the resonance position the amplitude is 
dominated by $\Theta^{+}$ terms with negligible
interference effects with background terms.  
Second, the theoretical situation for computing 
the background process is not well under control.  
Hence our estimation given here is a crude one, but will 
provide a guide line for the study of the relevant reactions.  

Here we concentrate on the process of the neutron target
exclusively.  
In this particular process we investigate in rather detail 
theoretical aspects of this reaction, since so far 
there are ambiguities that we can not settle, associated with 
reaction mechanism, interactions and $\Theta^+$ structure.  
Nevertheless, we attempt to draw some conclusions 
within acceptable approximations and ranges of unknown 
parameters.  
Furthermore, we consider a possibility to get information 
on some properties of $\Theta^+$ from this reaction.  
In particular, 
we consider the $\Theta^+$ production for two
possible cases of positive and negative parities of $\Theta^+$. 
The spin
and isospin are assumed to be $J=1/2$ and $I=0$. 
Motivated by theoretical interests, we perform calculations using
effective lagrangians of sigma models of linear 
(with a pseudoscalar coupling, PS) 
and non-linear 
(with a pseudovector coupling, PV)
types. 
Equivalence of the two methods in a certain case is used as a 
consistency check of 
our calculations.  
For quantitative discussions, we employ a hybrid model (HM) 
combining the 
PS and PV schemes with a suitable weight depending on the energy.  
We do not include heavier mesons such as scalar and vector mesons, 
since we do not know their interactions with $\Theta^+$.  
They may become important in actual processes.  
In  this sense, our model lagrangians are minimal.  

In a recent preprint, Liu and Ko studied various reactions
producing $\Theta^+$, including photon and hadron induced 
processes~\cite{Liu:2003rh}.  
For a photon induced process, they studied the case of a proton target, 
where they have assumed that $\Theta^+$ has positive parity.

This paper is organized as follow. In section II, we formulate
effective lagrangians of sigma model type including the kaon and
$\Theta^+$. 
Scattering amplitudes are then computed at the tree level.  
Results for numerical calculations are
presented in section III, where various cross sections are shown
for cases of positive and negative parity of $\Theta^+$. 
In the final section IV, we discuss results and summarize 
the present work.

\section{Effective lagrangians and amplitudes}

We consider interaction terms for the calculation of the 
reaction $\gamma n \to K^{-} \Theta^+$ at the tree level 
(in literatures, these terms are often referred to as the 
Born terms).  
In many cases, we show the formulae for the positive parity 
$\Theta^+$.  
The interactions with the negative parity $\Theta^+$ 
may be obtained by removing $\gamma_5$ in the following 
equations such as (1) and (2).  
The relevant interaction lagrangians 
(for the positive parity $\Theta^+$) satisfying gauge and chiral
symmetries are given by 
\bee
\mathcal{L}_{n\Theta K}
&=&
ig\bar{\Theta}\gamma_{5}K^{+}n,
\label{nTK_PS}
\; \; \; \; \; \; \; \; \; \; \; \; \; \; \; \; \; \; \; ({\rm PS}) \, ,  
\\
\mathcal{L}_{n\Theta K}
&=&
-\frac{g^{\ast}_{A}}{2f_{\pi}}
\bar{\Theta}\gamma_{\mu}\gamma_{5}\partial^{\mu}K^{+}n,
\label{nTK_PV}
\; \; \; \; \; ({\rm PV}) \, ,  
\\
\mathcal{L}_{\gamma KK}
&=&
ie\left\{ 
(\partial^{\mu}K^{-})K^{+}-(\partial^{\mu}K^{+})K^{-}
\right\}
A_{\mu},
\label{gKK}\\
\mathcal{L}_{\gamma nn}
&=&
-i\frac{e\kappa_{n}}{2M_{n}}\bar{n}\sigma_{\mu\nu}k^{\nu}n A^{\mu},
\label{gnn}\\
\mathcal{L}_{\gamma\Theta \Theta}
&=&
-e\bar{\Theta}\left(\gamma_{\mu}
+i\frac{\kappa_{\Theta}}{2M_{\Theta}}
\sigma_{\mu\nu}k^{\nu}\right)\Theta \, A^{\mu}\, .
\label{gTT}
\eee 
\normalsize
In these equations, $\Theta$, $n$ and $K$ denote the fields of
$\Theta^{+}$, the neutron and the kaon, respectively. 
In Eqs.~(\ref{nTK_PS}) and (\ref{nTK_PV}), $g$,
$g^{\ast}_{A}$ and $f_{\pi}$ are the coupling constant of 
$n \Theta^{+} K$ in PS, that in PV and the pion decay constant being
set $93.2$  MeV.  
The strengths of $g$ and $g^{\ast}_{A}$ will be determined later.  

In Eqs.~(\ref{gKK})-(\ref{gTT}), $e$ is a unit charge.  
In Eq.~(\ref{gnn}), $\kappa_n$ is the anomalous magnetic moment of the 
neutron (in units of nucleon magneton) and $k^\mu$ the photon momentum. 
In Eq.~(\ref{gTT}), we have introduced the anomalous magnetic
moment of $\Theta^{+}$, $\kappa_{\Theta}$ (in units of $\Theta$ magneton). 
We do not know its value, and therefore, treat it as a parameter in the
following calculations.   
For masses, we take $M_n = 939$ MeV and 
$M_\Theta = 1530$ MeV~\cite{Diakonov:1997mm}.  

\begin{figure}[p]
\begin{tabular}{ccc}
\resizebox{13cm}{4cm}{\includegraphics{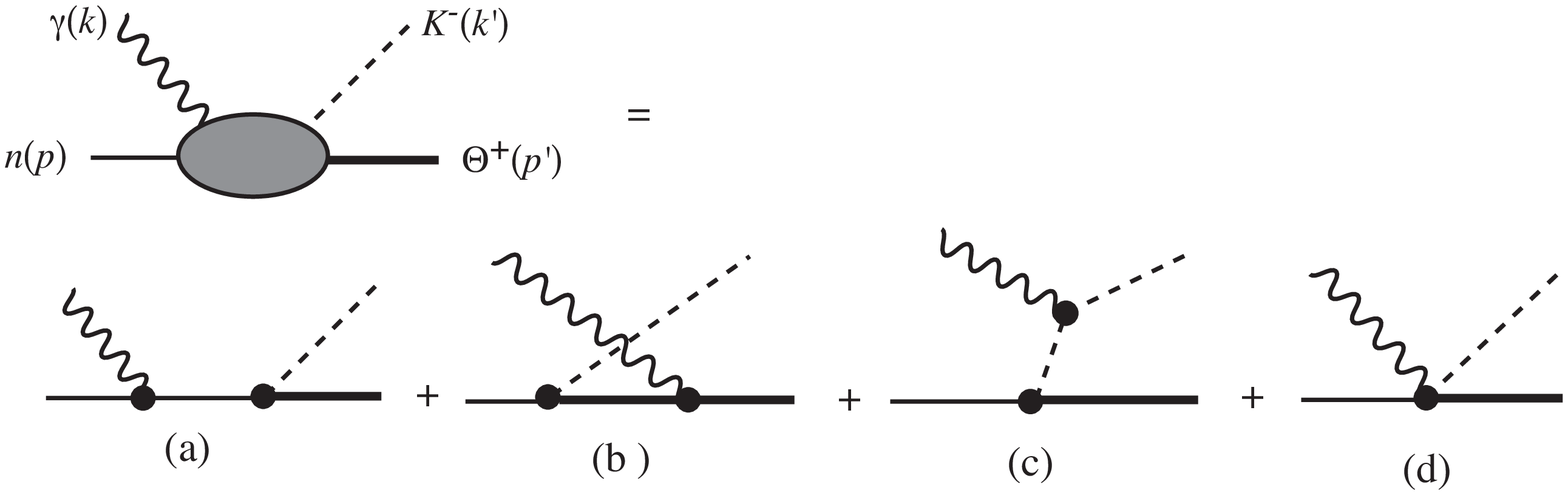}}
\end{tabular}
\caption{\small The tree diagrams for the 
   $\gamma n \to K^{-} \Theta^+$.
   (a) the direct Born Diagram (s-channel), 
   (b) the exchange Born diagram (u-channel), 
   (c) the meson exchange diagram (t-channel), and 
   (d) the contact (Kroll-Rudermann $\equiv$ KR) term. }
\label{FD}
\end{figure}

In the linear model of PS,  we use Eqs.~(\ref{nTK_PS}),
(\ref{gKK})-(\ref{gTT}), and compute the three terms as shown in 
Figs.~\ref{FD} (a) -- (c).  
The resulting amplitude satisfies gauge invariance.  
In contrast, in the non-linear model of PV, in addition 
to the three terms, we need the contact term as shown in 
Fig.~\ref{FD} (d), which is obtained by the replacement
$\partial_{\mu}\rightarrow \partial_{\mu}+i e A_{\mu}$ in the lagrangian Eq.~(\ref{nTK_PV}).

Defining  $S={\bf 1}+iT$ and 
$
\langle p',k'|iT|p,k\rangle
=
i (2\pi)^{4}\delta^{(4)}(p+k-p'-k')
\mathcal{M}
$, 
in PS the gauge invariant amplitudes become
\bee
i\mathcal{M}_{\rm s}
&=& 
eg\frac{\kappa_{n}}{4M_{n}}\bar{u}(p')\gamma_{5}
\frac{\Slash{p}+\Slash{k}+M_{n}}
{(p+k)^{2}-M^{2}_{n}}
(\Slash{\epsilon}\Slash{k}-\Slash{k}\Slash{\epsilon})u(p),
\label{M2}\\
i\mathcal{M}_{\rm u}
&=&
-eg\bar{u}(p')\Slash{\epsilon}
\frac{\Slash{p'}-\Slash{k}+M_{\Theta}}
{(p'-k)^{2}-M^{2}_{\Theta}}
\gamma_{5}u(p) 
\nn \\
&+& eg\frac{\kappa_{\Theta}}{4M_{\Theta}}
\bar{u}(\Slash{\epsilon}\Slash{k}-\Slash{k}\Slash{\epsilon})
\frac{\Slash{p'}-\Slash{k}+M_{\Theta}}
{(p'-k)^{2}-M^{2}_{\Theta}}\gamma_{5}u(p),
\label{M4}\\
i\mathcal{M}_{\rm t}
&=&
eg\bar{u}(p')
\frac{\gamma_{5}}{(k-k')^{2}-m^{2}_{K^{+}}}
u(p)(2k'\cdot\epsilon-k\cdot \epsilon),\label{M3}
\eee
where $\bar{u}$ and $u$ are the Dirac spinors of $\Theta^+$ and the
neutron. $p$, $p'$, $k$ and $k'$ are the four-momenta of  
the neutron, $\Theta^{+}$, photon and $K^{-}$, respectively. 
Subscripts s, u and t 
indicate the  channels corresponding to Figs.~\ref{FD}
(a), (b) and (c).

In the nonlinear framework of PV, 
the four diagrams of Figs.~\ref{FD} (a) -- (d) must be 
computed.  
The first three terms are given by the substitution 
in the result of PS, 
$\gamma_{5}\rightarrow \gamma_{5}\Slash{k}'$ 
in Eqs.~(\ref{M2}) and (\ref{M4}), 
$\gamma_{5} \rightarrow
\gamma_{5}(\Slash{k}'-\Slash{k})$ in Eq.~(\ref{M3}).  
The Kroll-Rudermann term is computed separately as 
\be
i\mathcal{M}_{\rm KR,}
=
-e\frac{g^{\ast}_{A}}{2f_{\pi}}\bar{u}(p')\gamma_{5}\Slash{\epsilon}u(p)
\label{M1}.
\ee

Now we determine the
coupling constants, $g^{\ast}_{A}$ and 
$g$ from the decay width of $\Theta^+$.   
We use a formula 
%
\bee
\Gamma=\frac{1}{M_{i}}\frac{1}{4\pi}
\frac{|\vec{P}_{f}|}{E_{cm}}
\sum_{\rm final \; state}|\mathcal{M}|^{2}\, , 
\eee
%
where $\vec{P}_{f}$, $E_{cm}$ and $M_{i}$ are the relative 
three momentum of the final state,
the total energy in the center of mass (CM) energy frame and 
the mass of the initial particle, respectively. 
We use
Eqs.~(\ref{nTK_PS}), (\ref{nTK_PV}) and $\Gamma=15$ MeV from 
chiral soliton 
model calculations. 
Then, we obtain
$g^{\ast}_{A}=0.306$ for PV and $g=4.05$ for PS. If $\Theta^+$ has
negative parity,  
then we find 
$g^{\ast}_{A}=0.171$ and $g=0.542$. 
It is interesting that these coupling constants 
satisfy the (extended) Goldberger-Trieman 
relations 
\bee
\frac{g^{\ast}_{A}}{2f_{\pi}}
&=&
\frac{g}{M_{n}+M_{\Theta}} \, , \; \; \; \; 
P(\Theta^+) = + \, , \\
\frac{g^{\ast}_{A}}{2f_{\pi}}
&=&
\frac{g}{M_{\Theta}-M_{n} } \, , \; \; \; \; 
P(\Theta^+) = - \, .
\eee
Because of these GT relations, the amplitudes 
computed in the PS and PV become 
equivalent, when $\kappa_n = \kappa_{\Theta} =0$.  
This provides a good consistency check of our 
numerical calculations.  

In reactions such as pion and kaon photoproductions, 
it is known that only the tree (or Born) diagrams are not 
sufficient to reproduce experimental data.  
In fact, as the energy is increased the cross 
sections of the Born terms rise too rapidly.  
In the hybrid model (HM) the linear and non-linear 
models are combined under the assumption that at low energies 
the non-linear model works well while at higher energies, 
the linear one does better.   
As proposed in Refs.~\cite{Drechsel:1998hk,Katagiri:bt}, we consider the
following HM amplitude
\bee
i \mathcal{M}_{\rm HM} = i \mathcal{M}_{\rm PS}
+i\Delta\mathcal{M}\cdot F_{\rm HM}\,  ,\; \; \; \; 
\Delta\mathcal{M} \equiv \mathcal{M}_{\rm PV} - \mathcal{M}_{\rm PS}
\label{HM}
\eee
where 
$
F_{\rm HM} = 
\Lambda_{\rm HM}^{2}/
(\Lambda_{\rm HM}^{2}
+ |\vec{k}'|^{2})$ and $\Lambda_{\rm HM}$ is the mixing parameter
depending on the kaon  
momentum $\vec{k}'$.  
Obviously the amplitude of HM satisfies the gauge invariance.  
For the mixing parameter, we employ 
$\Lambda_{\rm HM}=450$ MeV~\cite{Drechsel:1998hk}.  
Now we find for the difference between the PS and PV amplitudes 
\bee
i\Delta\mathcal{M} = 
e \frac{g^{\ast}_{A}}{2f_{\pi}}
\left( \frac{\kappa_{\Theta}}{2M_{\Theta}}
+ \frac{\kappa_{n}}{2M_{n}}
\right) 
\bar{u}(p')\gamma_{5}\Slash{\epsilon}\Slash{k}u(p) \, .
\label{delta}
\eee
Since $\kappa_{n}=-1.91$, for a positive value of
$\kappa_{\Theta}$ the difference 
$\Delta\mathcal{M}$ 
is suppressed, while for a negative value enhanced.  
As discussed in the next section, 
the effect of the mixing of the PS and PV in the HM  
is to make the energy dependence of the cross sections 
milder.  

\begin{figure}[ht]
\begin{tabular}{ccc}
\resizebox{5.5cm}{4cm}{\includegraphics{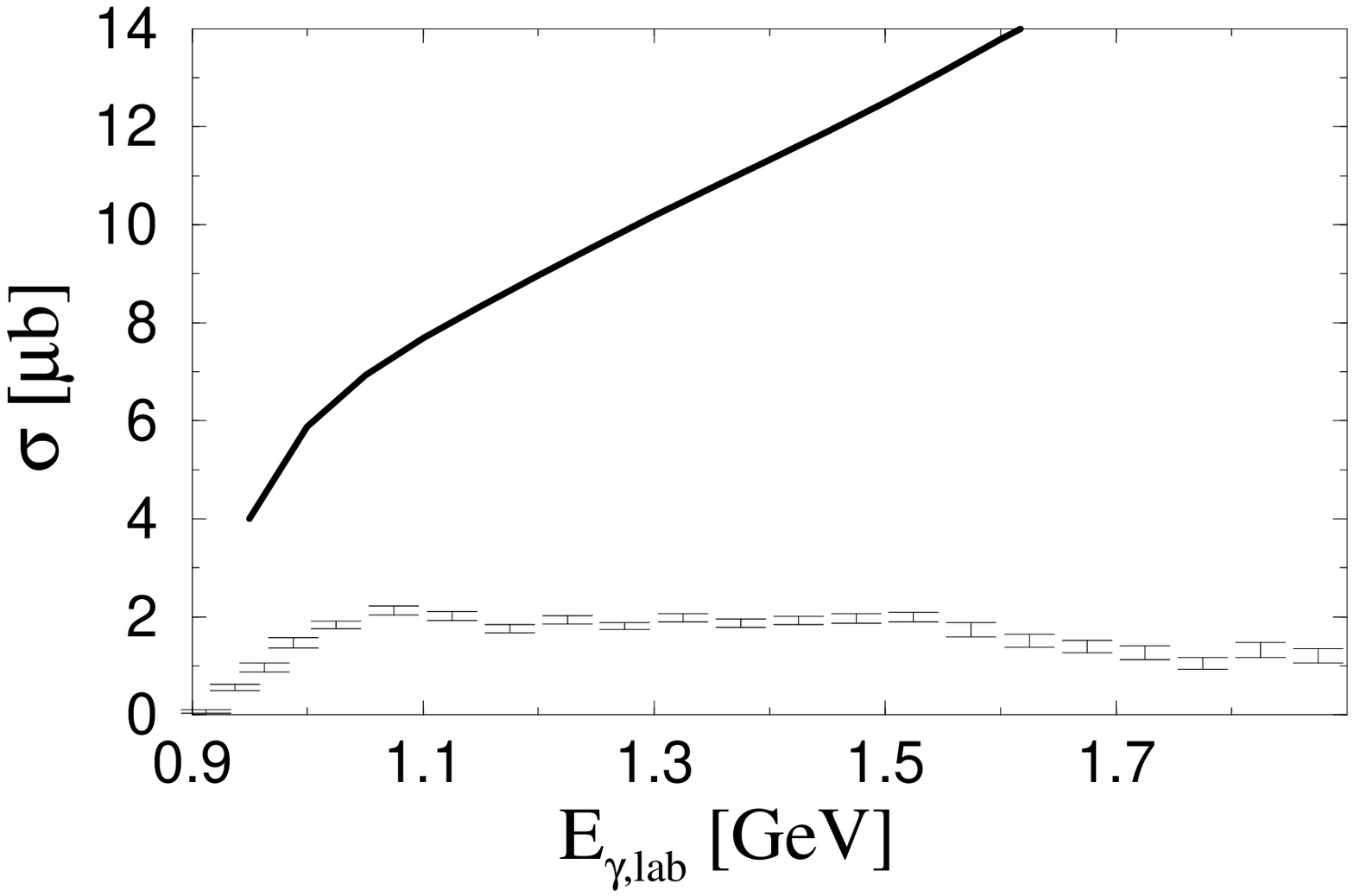}}
\resizebox{5.5cm}{4cm}{\includegraphics{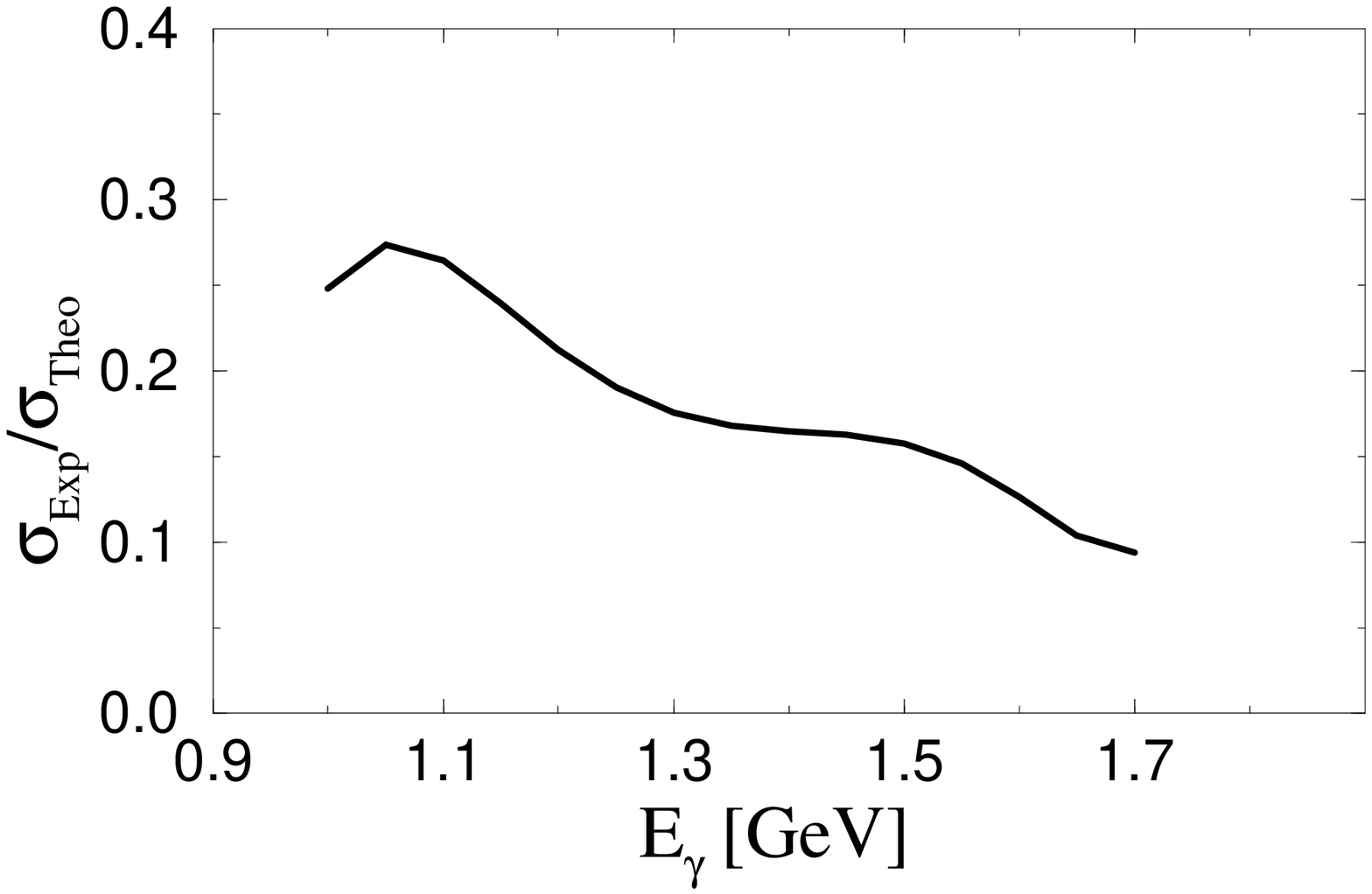}}
\end{tabular}
\caption{Left panel: the total cross sections
of $p(\gamma,K^{+})\Lambda$~\cite{Janssen:2001wk}. The theoretical
result (solid line) was calculated by the Born terms using an SU(3)
value of $g_{KN\Lambda}$. The experimental data are taken from
Ref.~\cite{Tran:qw}. Right panel: the ratio of the experimental data
and the theoretical calculation.} 
\label{GPKL}
\end{figure}

There is another factor we need to consider, which is the 
effect of the form factor.  
The status of theoretical treatment is still controversial
due to subtleties of maintaining gauge invariance.  
However, 
the necessity of the form factor seems obvious from physics point 
of view as it should reflect the size (or internal structure) 
of hadrons.  
For instance, the Born terms in the $K^+\Lambda$
photoproduction~\cite{Janssen:2001wk} 
produces too large total cross sections 
as shown in  Fig.~\ref{GPKL}. From the comparison in the theoretical
and experimental  
cross sections of $K\Lambda$
photoproduction, we may naturally expect a suppression also for 
the present reaction of $K^-\Theta^+$ production.  
More details will be discussed in the next section.  

%
\section{Numerical results}

We present the total and differential cross sections
for the cases of positive and negative parity 
$\Theta^+$ using the lagrangians of PS, PV and HM. 
present the results with using different anomalous magnetic moment of 
$\Theta^+$; $\kappa_{\Theta} = 1.79, 0.0$ and 
$-1.79$ in the units of $\Theta$ baryon.  
If we assume a particular structure of $\Theta^+$, it is 
possible to estimate the size of the magnetic moment.  
For instance, in a model of Jaffe-Wilczek where two $S$ (total spin)
$= 1$ 
$[ud]$ 
diquarks are in relative $p$-wave motion for positive 
parity $\Theta^+$, 
$\kappa_{\Theta} \sim -0.7$, while 
$\kappa_{\Theta} \sim - 0.2$ if they are in $s$-state
for negative parity $\Theta^+$
(in this case one of $ud$ diquaks has $S = 1$).  
If $\Theta^+$ would be a $KN$ bound state in a $p$-orbit for a positive 
parity $\Theta^+$, $\kappa_{\Theta} \sim - 0.4$, while 
$\kappa_{\Theta} \sim - 0.5$ if they are in $s$-state
for a negative parity $\Theta^+$. 
In all cases, $\kappa_\Theta$ takes a value around 
-0.5. These numbers may be compared with a recent evaluation in the
chiral soliton model which provided
$\kappa_{\Theta}=0.3$~\cite{Kim:2003ay}. In any event, we 
treat $\kappa_{\Theta}$  
as unknown parameter in the following calculations.  
As we have mentioned, the equivalence of the 
PS and PV amplitudes when $\kappa = 0$ has been  
used as a check of our numerical calculations.

\begin{figure}[ht]
\begin{tabular}{ccc}
\resizebox{5.5cm}{4cm}{\includegraphics{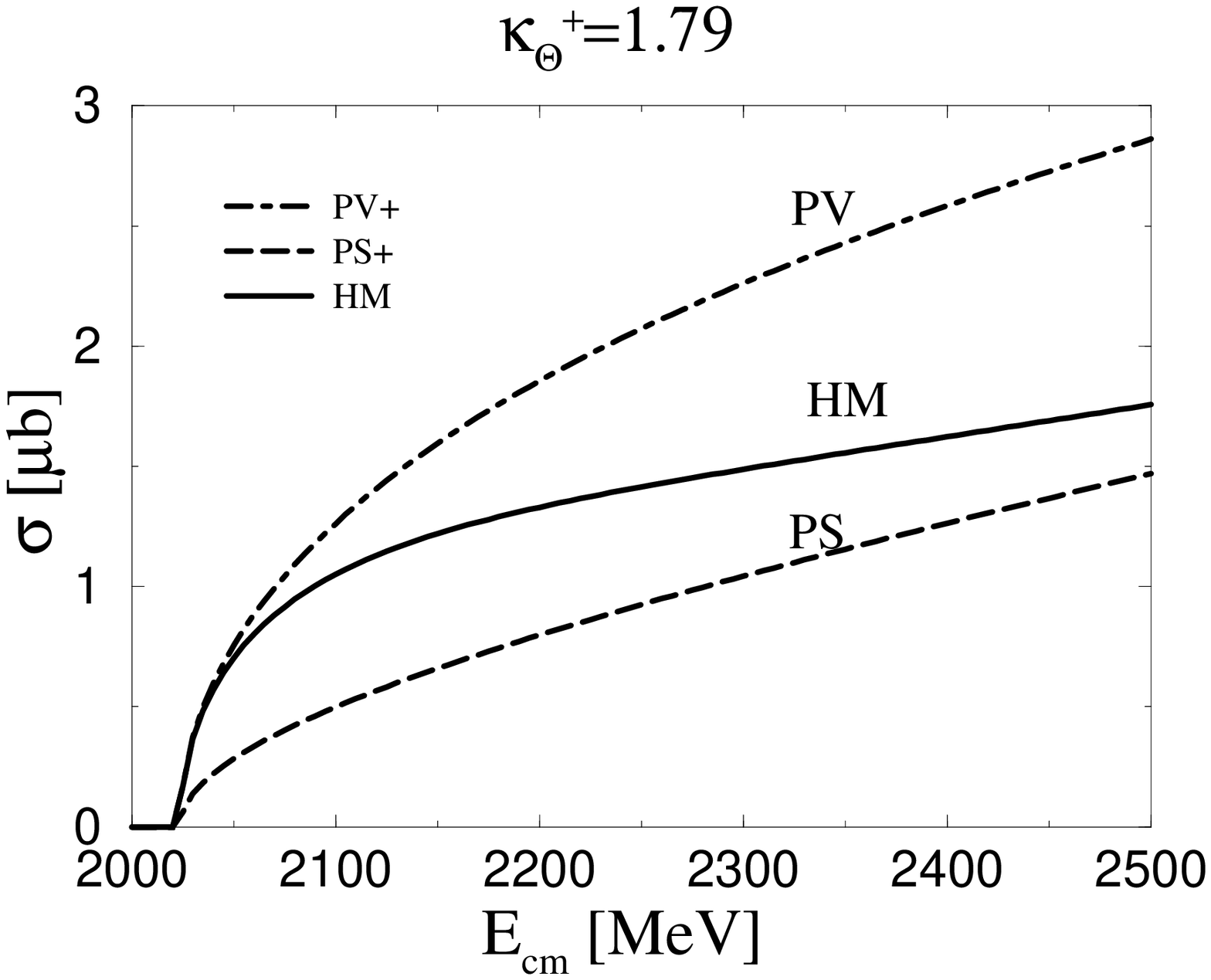}}
\resizebox{5.5cm}{4cm}{\includegraphics{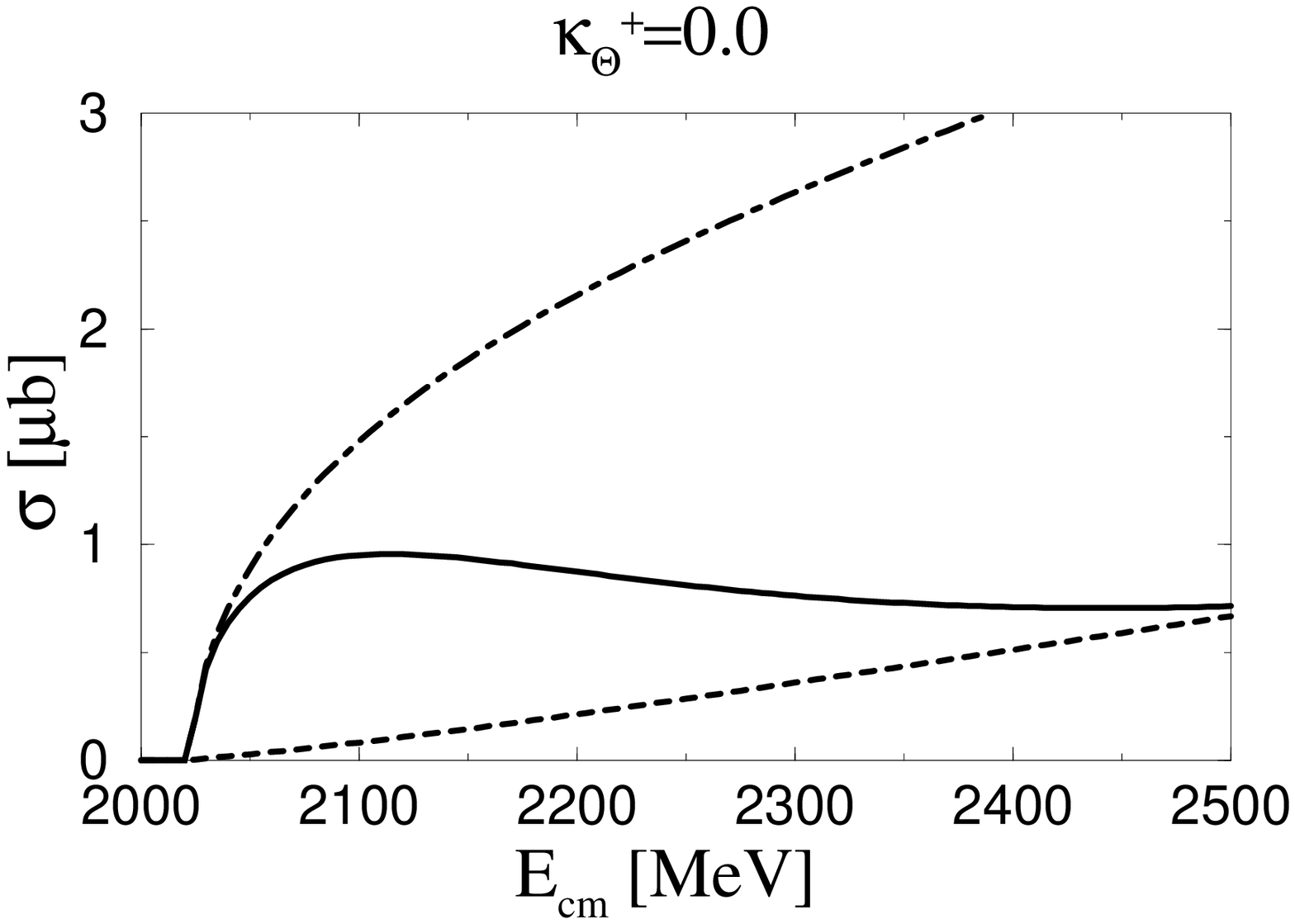}}
\resizebox{5.5cm}{4cm}{\includegraphics{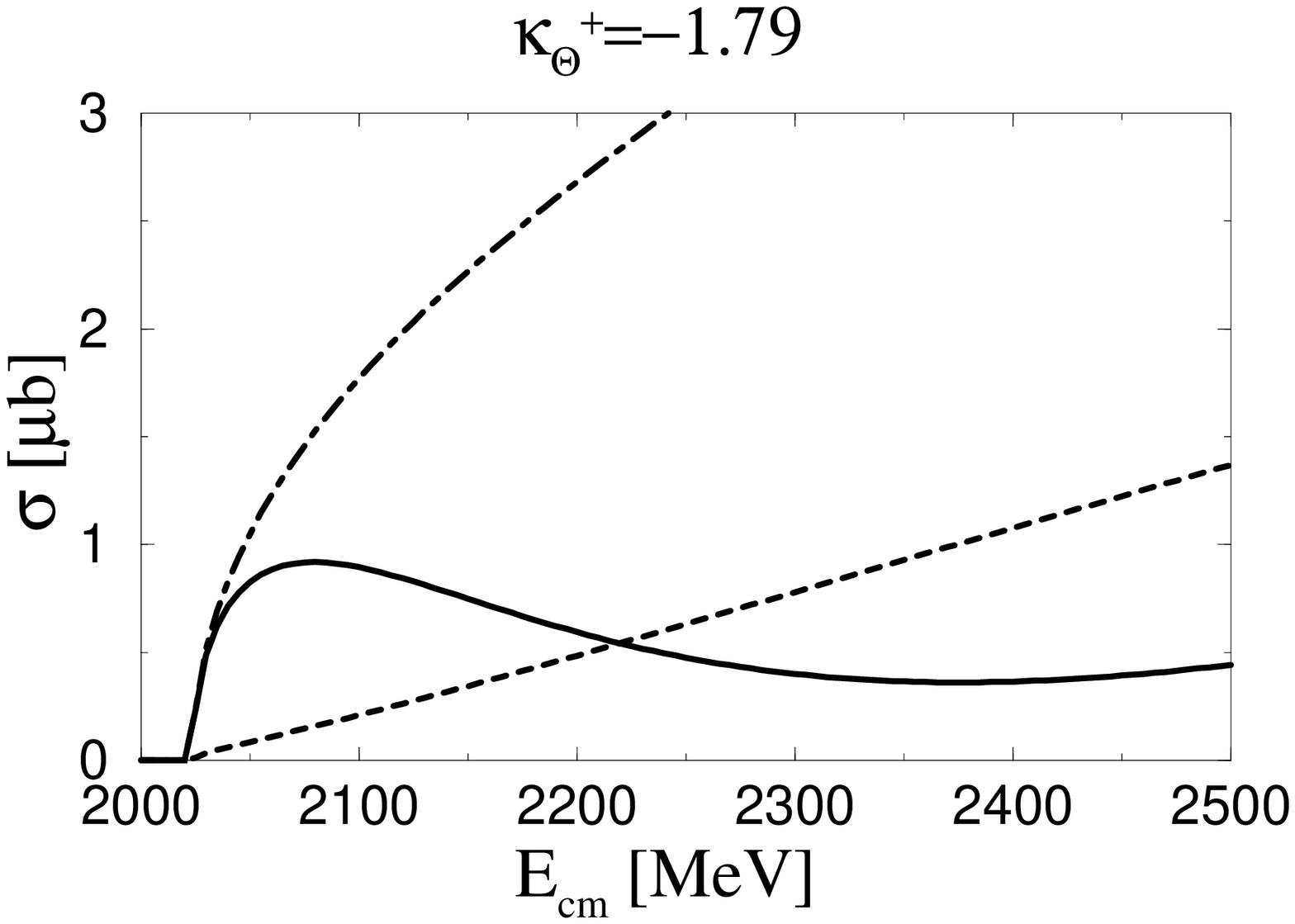}}
\end{tabular}
\caption{Total cross sections for three different $\kappa_{\Theta}$'s
for $J^P=1/2^{+}$. 
Solid, dotted and dot-dashed lines are for HM, PS and PV, respectively.} 
\label{TCS}
\end{figure}

In Figs.~\ref{TCS} the total cross sections are shown as 
functions of the total energy in the center of mass 
system $E_{CM}$.   
The results depend significantly on the choice of a lagrangian 
scheme and $\kappa_{\Theta}$. As $E_{CM}$ is increased, the cross
sections of PS and PV (and HM 
when $\kappa_{\Theta} = 1.79$) increase.  
Especially, the rate of the increase in the PV seems very large. 
However, such a behavior may not be realistic if we recall 
the result of $K\Lambda$ production as seen in Fig.~\ref{GPKL}
In fact, this has motivated us to employ the scheme of the hybrid model 
(HM).  
As shown in the figures, the energy dependence of the 
HM result is much milder, and the total cross section is 
suppressed at large $E_{CM}$.    
Because of this, in the following  discussions we often 
refer to the results of the HM when we address actual numbers.  
The total cross section in HM averaged over the energy range 
$E_{CM} \lsim 2.5$ GeV is thus about 
1.34 $\mu$b (for $\kappa_{\Theta} = + 1.79$), 
0.78 $\mu$b (for $\kappa_{\Theta} = 0$), 
0.54 $\mu$b (for $\kappa_{\Theta} = - 1.79$).

Yet we need to consider the effect of form factors associated with  
extended structure of hadrons, as we have discussed in the previous 
section in Fig~\ref{GPKL}.  
Indeed, the energy dependence of the PS cross section for the $K\Theta$ 
production is similar to that of 
$K \Lambda$ production when $\kappa_\Theta = + 1.79$ is used.  
The difference in the magnitude may be  understood by the 
difference in the coupling constants, 
$g_{KN\Theta} \sim 
(1/2.5)  g_{KN\Lambda}({\rm SU(3) \; value})\sim 4$.  
As shown in the right panel of Fig.~\ref{GPKL}, 
the average rate of the suppression is about 
18 \% 
in the energy range of 0.9 GeV $\lsim E_{\gamma} \lsim 2.0$ GeV
($E_{threshold} \lsim E_{CM} \lsim 2.15$ GeV).  
If we expect a similar suppression factor also 
for the $K\Theta$ production,  
then a typical total cross section for  
$\gamma n \to K^- \Theta^+$ averaged over the 
energy range 
$E_{threshold} \lsim E_{CM} \lsim 2.5$ GeV 
would be about 
240 nb (for $\kappa_{\Theta} = + 1.79$), 
140 nb (for $\kappa_{\Theta} = 0$), 
100 nb (for $\kappa_{\Theta} = - 1.79$).  

\begin{figure}[ht]
\begin{tabular}{ccc}
\resizebox{5.5cm}{4cm}{\includegraphics{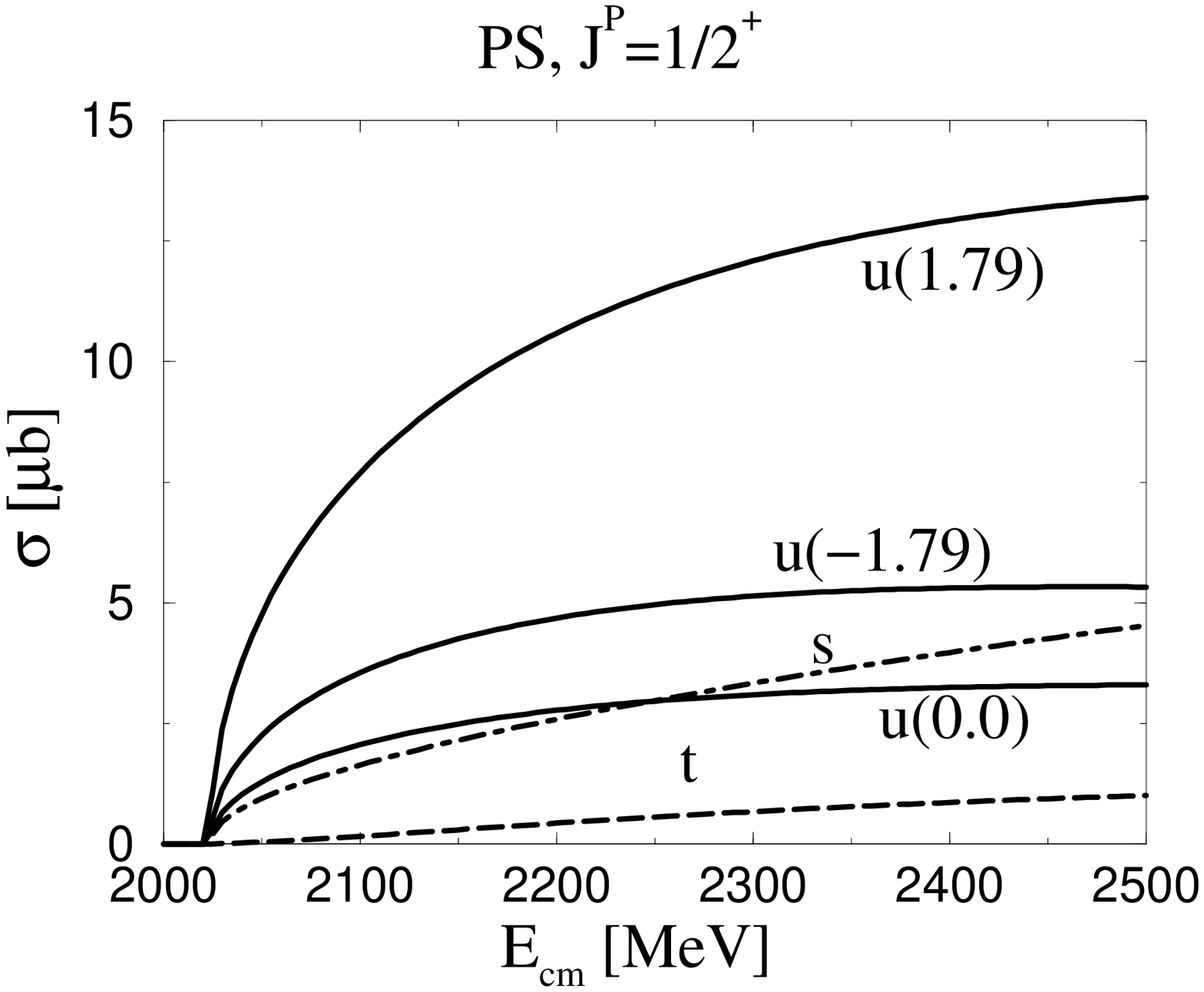}}
\resizebox{5.5cm}{4cm}{\includegraphics{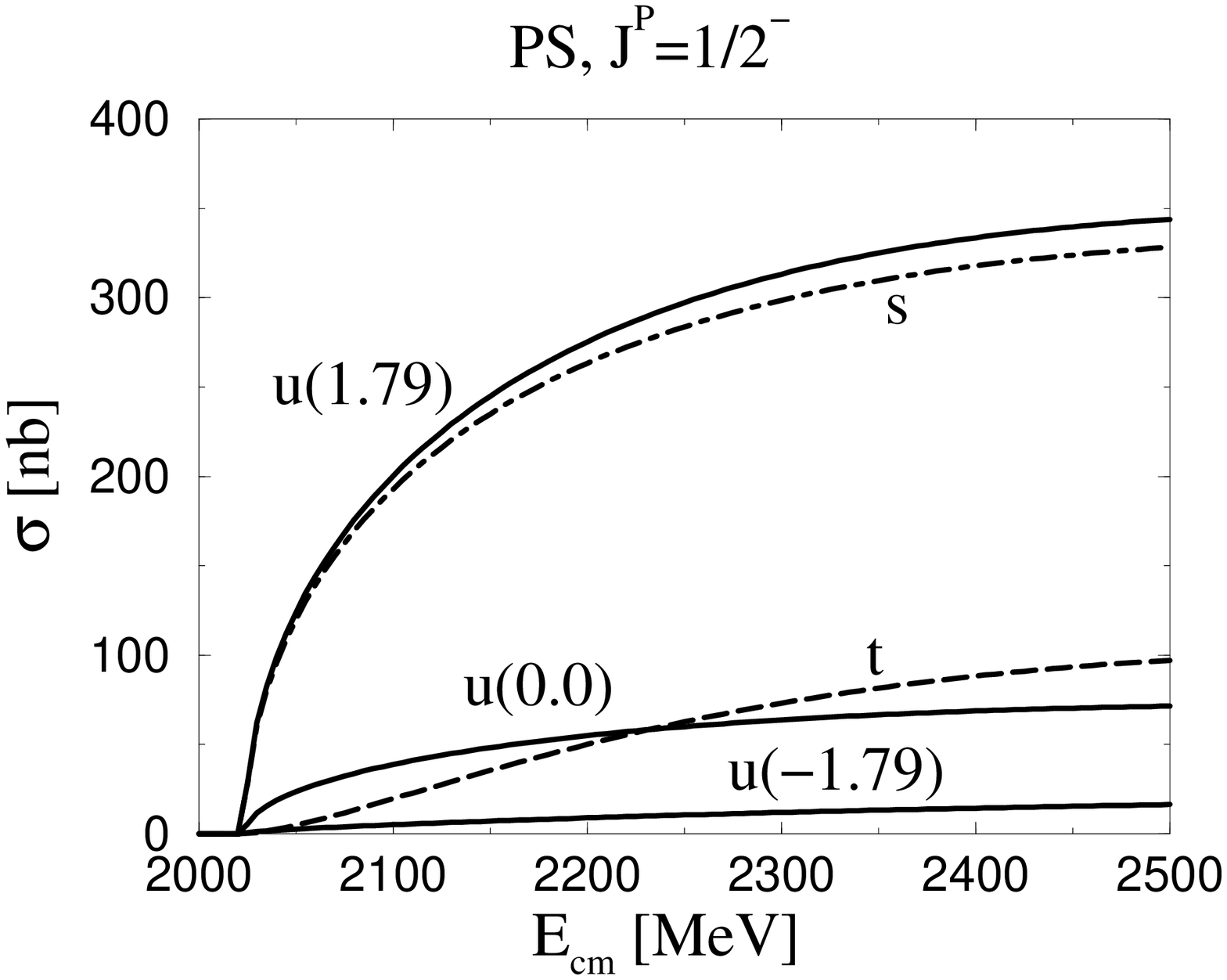}}
\end{tabular}
\begin{tabular}{ccc}
\resizebox{5.5cm}{4cm}{\includegraphics{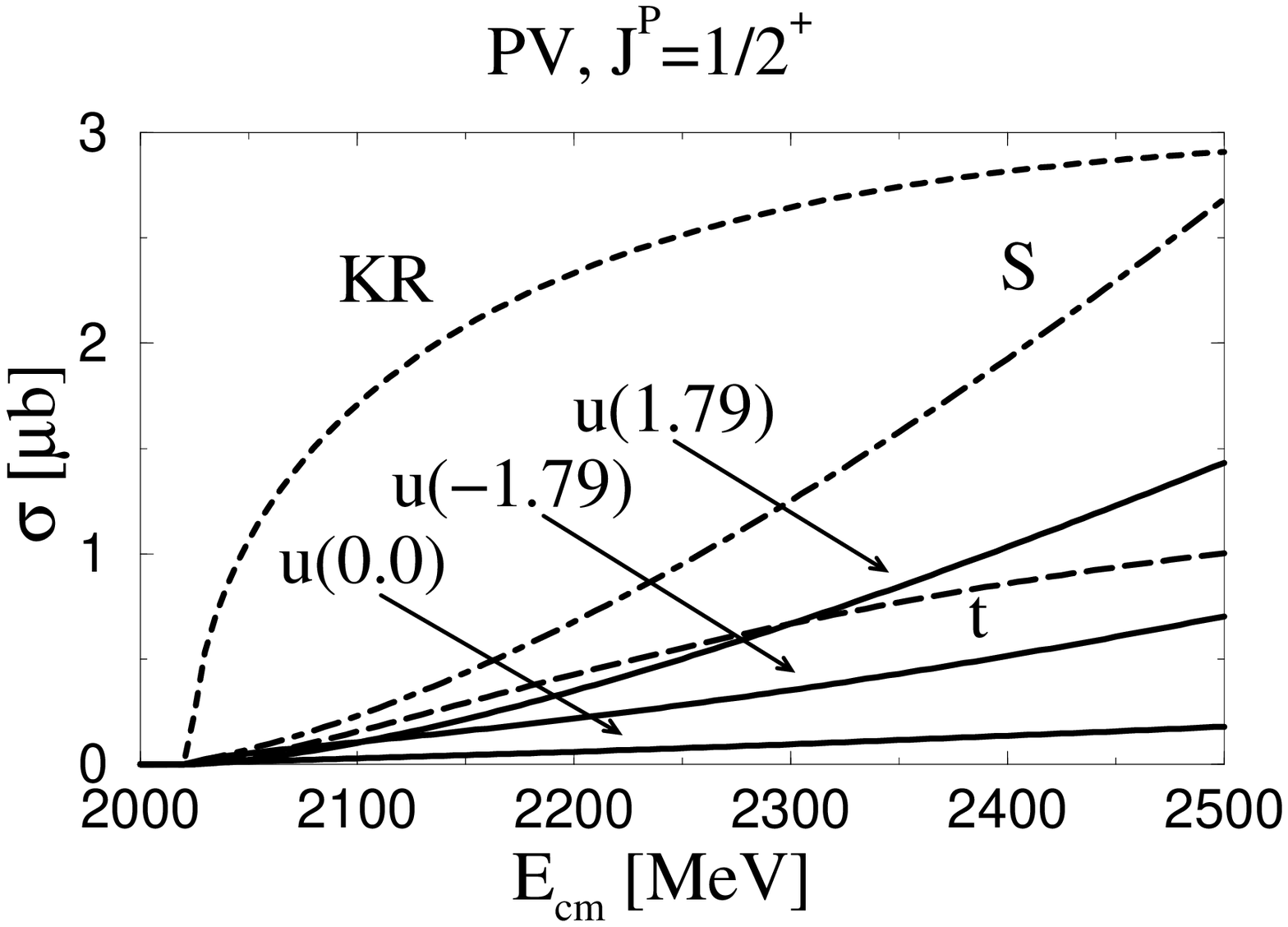}}
\resizebox{5.5cm}{4cm}{\includegraphics{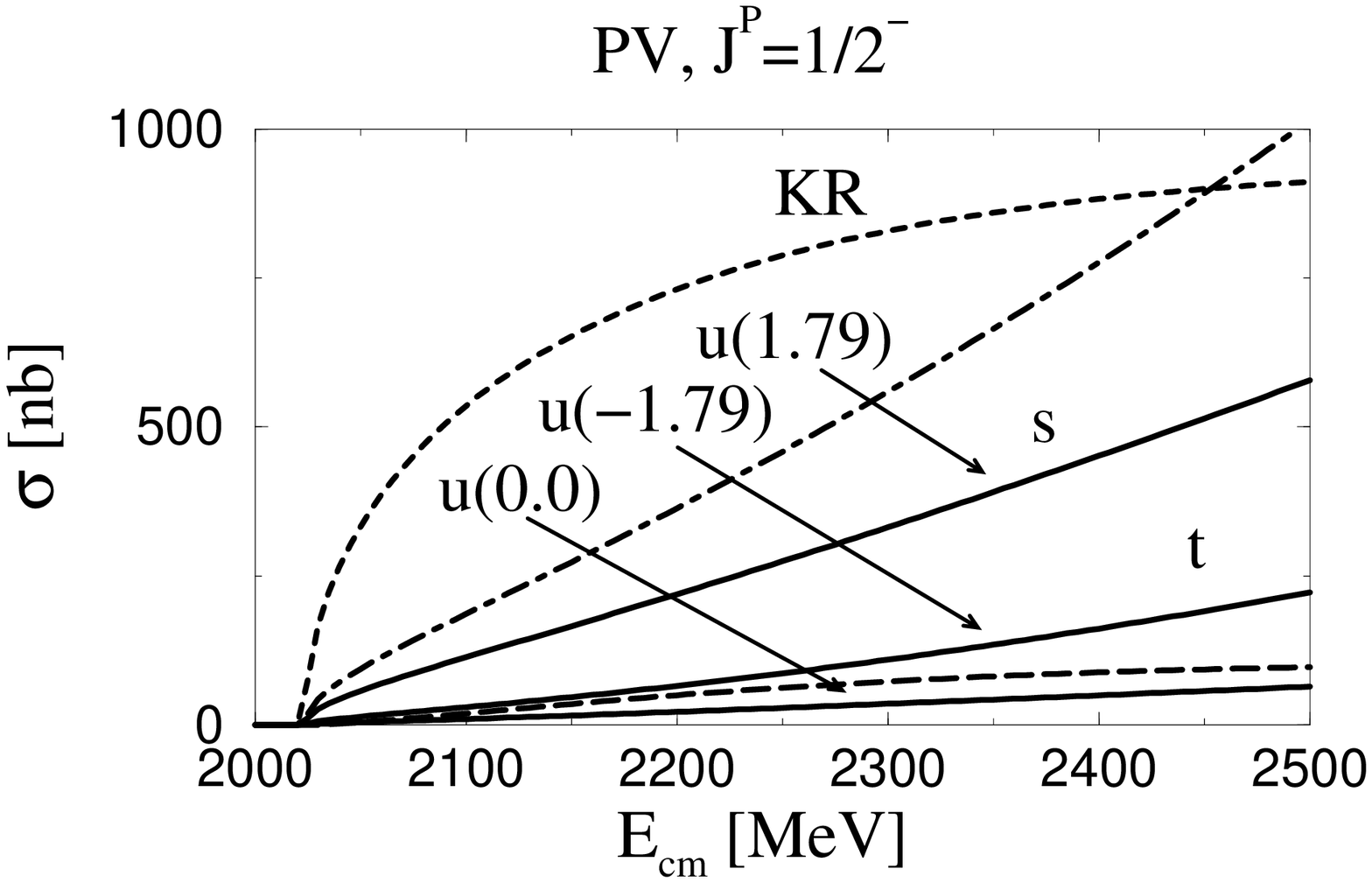}}
\end{tabular}
\caption{Contributions from various term in PS and PV schemes for positive
and negative parity. Dashed lines for t-channel, dot-dashed for
s-channel and solid for u-channel contributions. For PV, dotted line for the
contact term (KR).} 
\label{sut}
\end{figure}

At this point it would be useful to demonstrate contributions from 
each term of the diagrams.  
In Figs.\ref{sut}, we show contributions from the 
direct Born (s-channel), exchange Born (u-channel), 
kaon exchange (t-channel), and contact (KR) terms.  
The s, t and KR terms do not depend on $\kappa_\Theta$, while 
the u-channel contribution does as seen from Eqs.~(\ref{M2})-(\ref{M1}).  
In PS, the u and s-channel contributions are dominant, while 
in PV, the contact KR term dominates.  
The t-channel contribution is relatively small. 
Also from the figures, we observe that 
the threshold behavior of the s, u and KR terms are different from 
that of the t-channel meson exchange term.  
The former produces the $E_{0^+}$ amplitude
in S-wave of the final state $K \Theta^+$, while 
the latter is for higher multipoles.  

\begin{figure}[ht]
\begin{tabular}{ccc}
\resizebox{5.5cm}{4cm}{\includegraphics{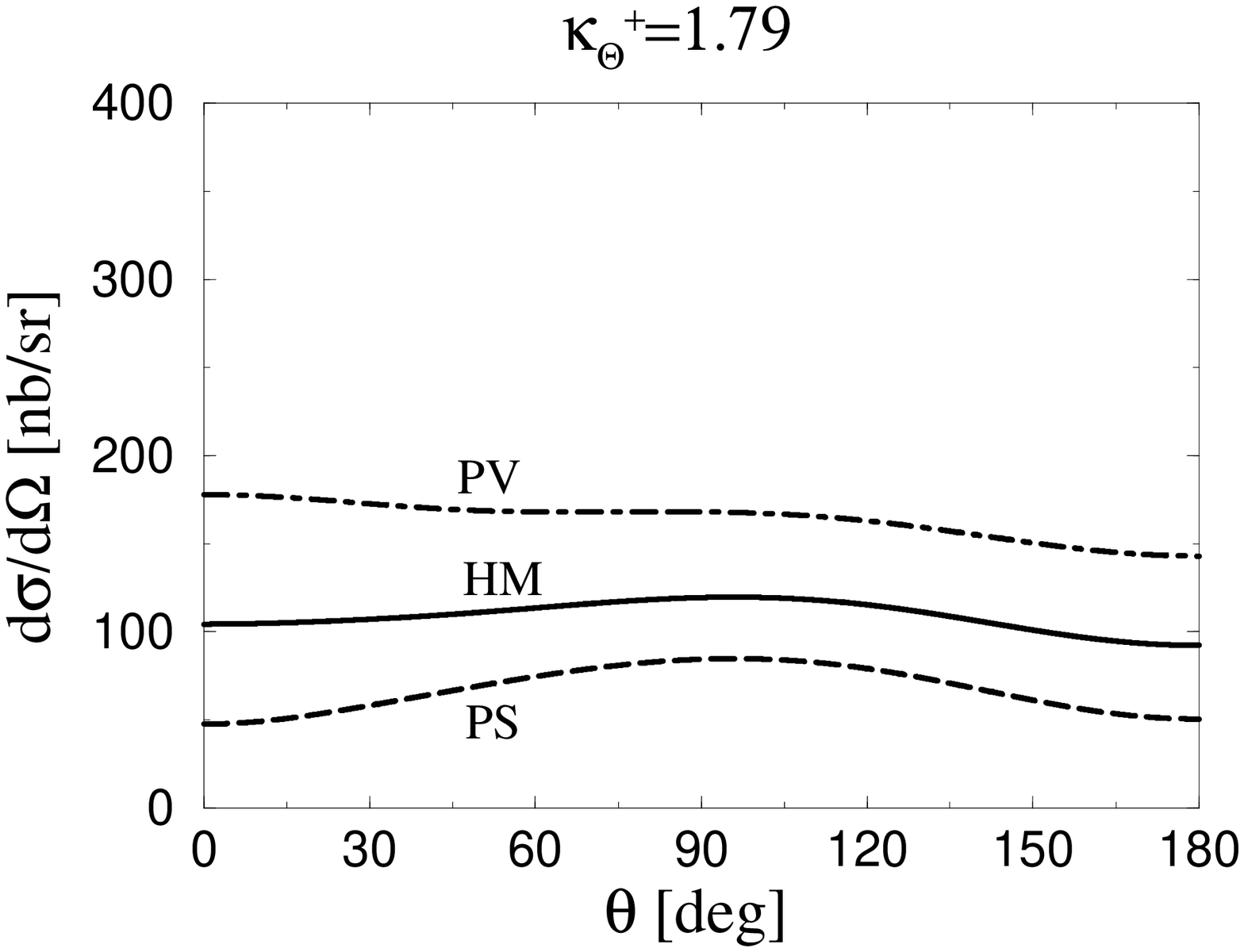}}
\resizebox{5.5cm}{4cm}{\includegraphics{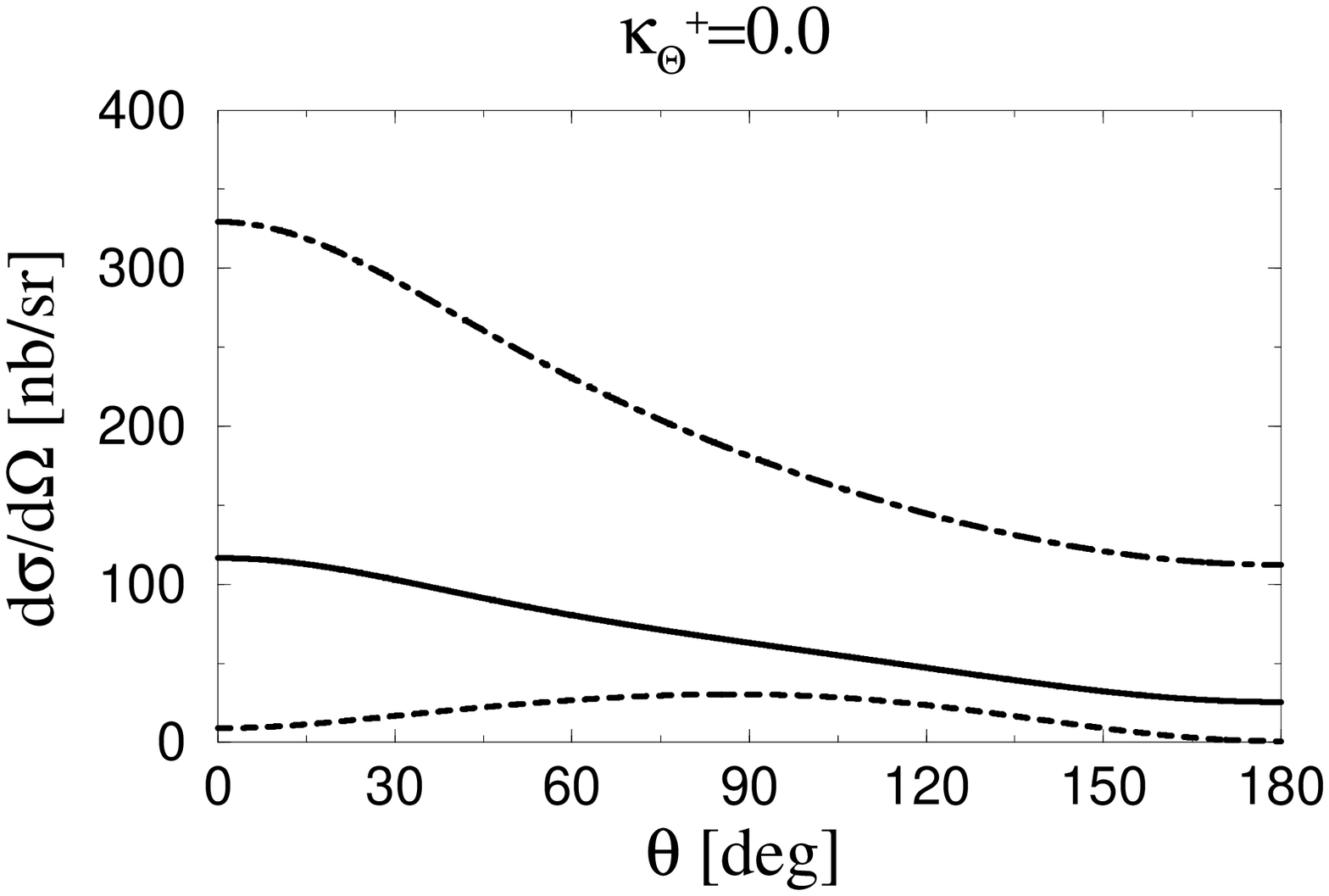}}
\resizebox{5.5cm}{4cm}{\includegraphics{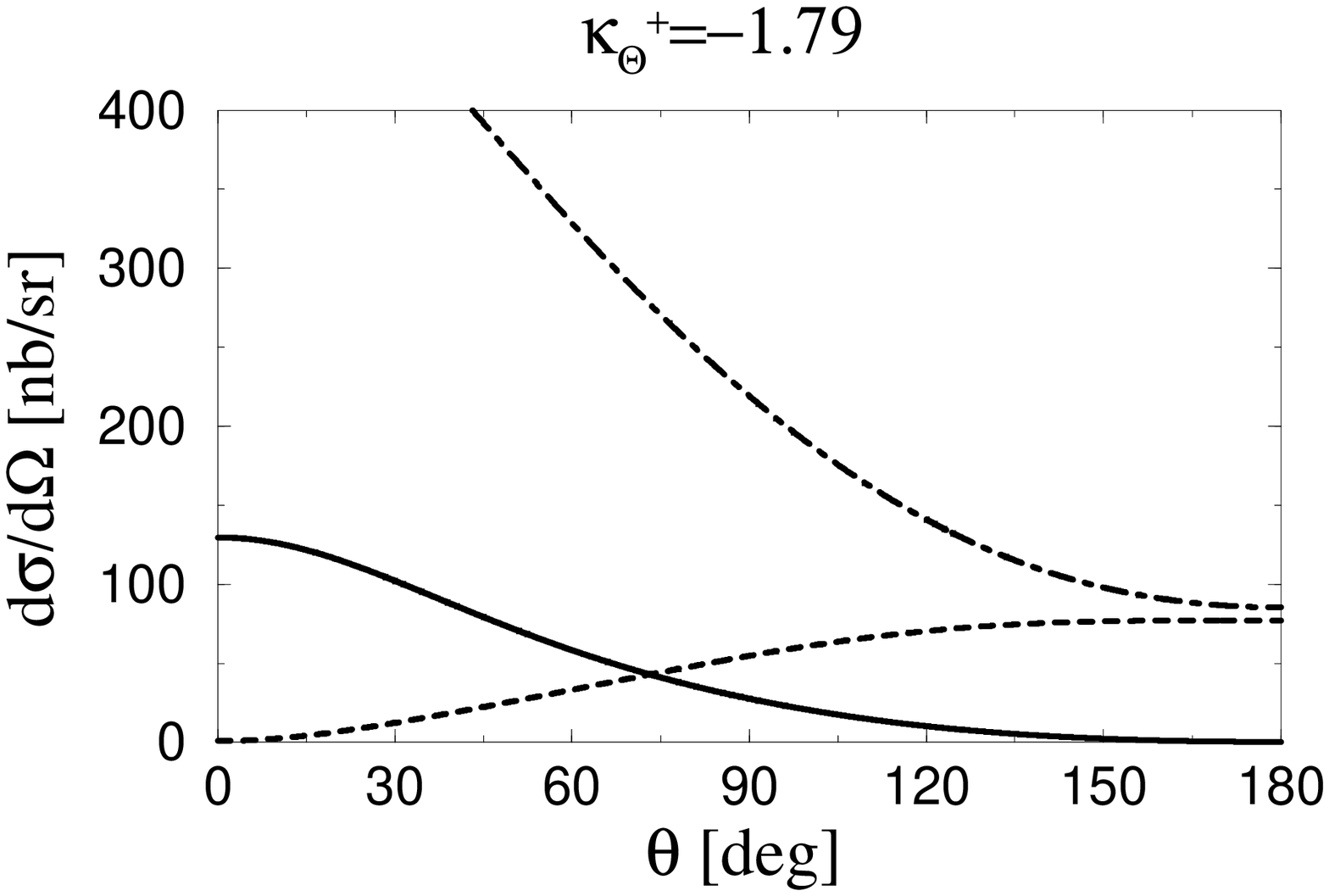}}
\end{tabular}
\caption{Differential  
cross sections for three different $\kappa_{\Theta}$'s
for $J^P=1/2^{+}$. Solid, dotted and dot-dashed lines are for HM,
PS and PV, respectively.}
\label{DCS}
\end{figure}

In Fig.\ref{DCS} differential cross
sections at a typical value $E_{CM} = 2.25$ GeV are shown 
for the same $\kappa_\Theta$ as for the total cross
sections. The angular dependence also vary significantly depending on the 
lagrangian scheme and on $\kappa_{\Theta}$.  
However, except for the PS case, the cross sections are 
forward peaking especially when a smaller $\kappa_{\Theta}$ is
employed. At this point we note that when $\kappa_\Theta\sim 0$, the
angular dependence has a forward peak for the cases of HM and PV. 

\begin{figure}[ht]
\begin{tabular}{ccc}
\resizebox{5.5cm}{4cm}{\includegraphics{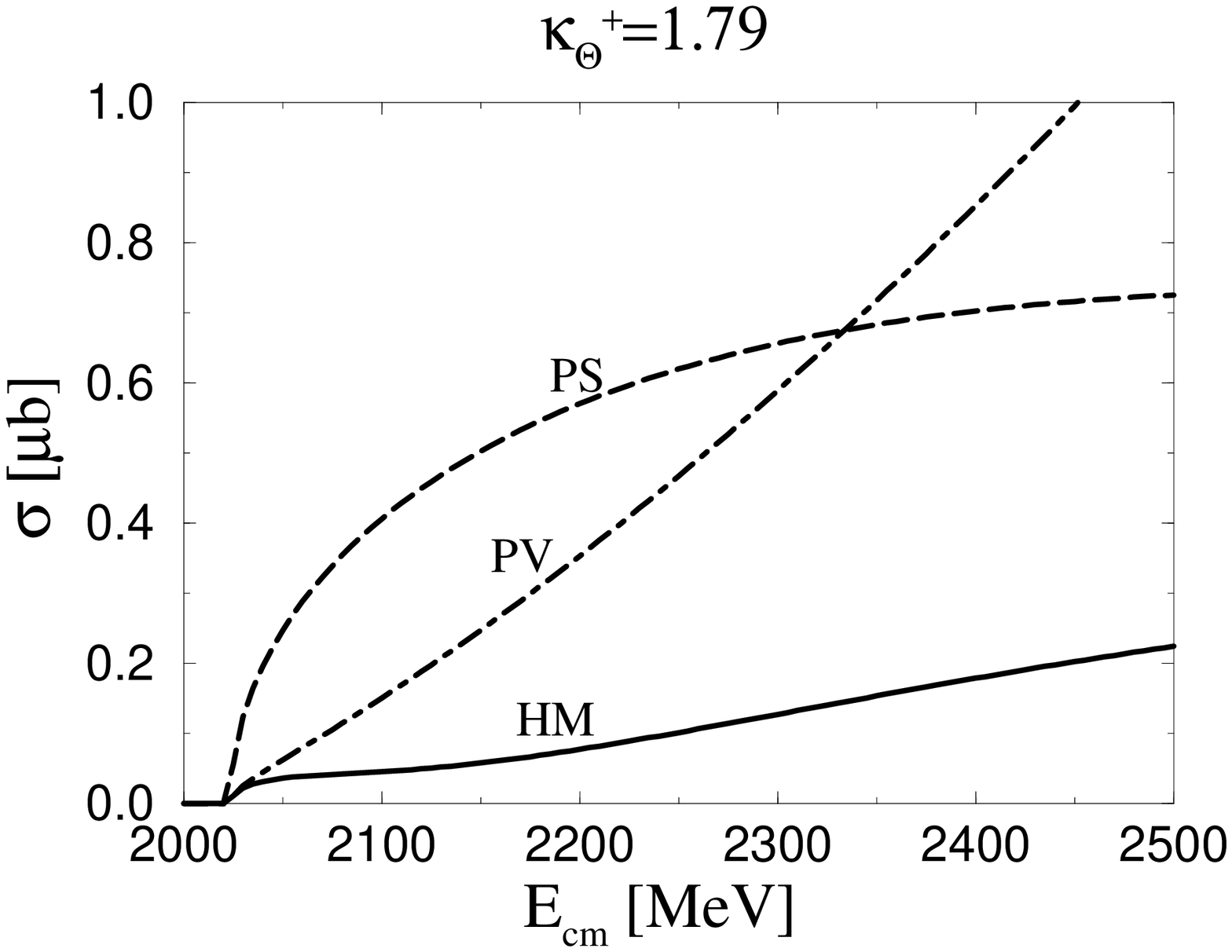}}
\resizebox{5.5cm}{4cm}{\includegraphics{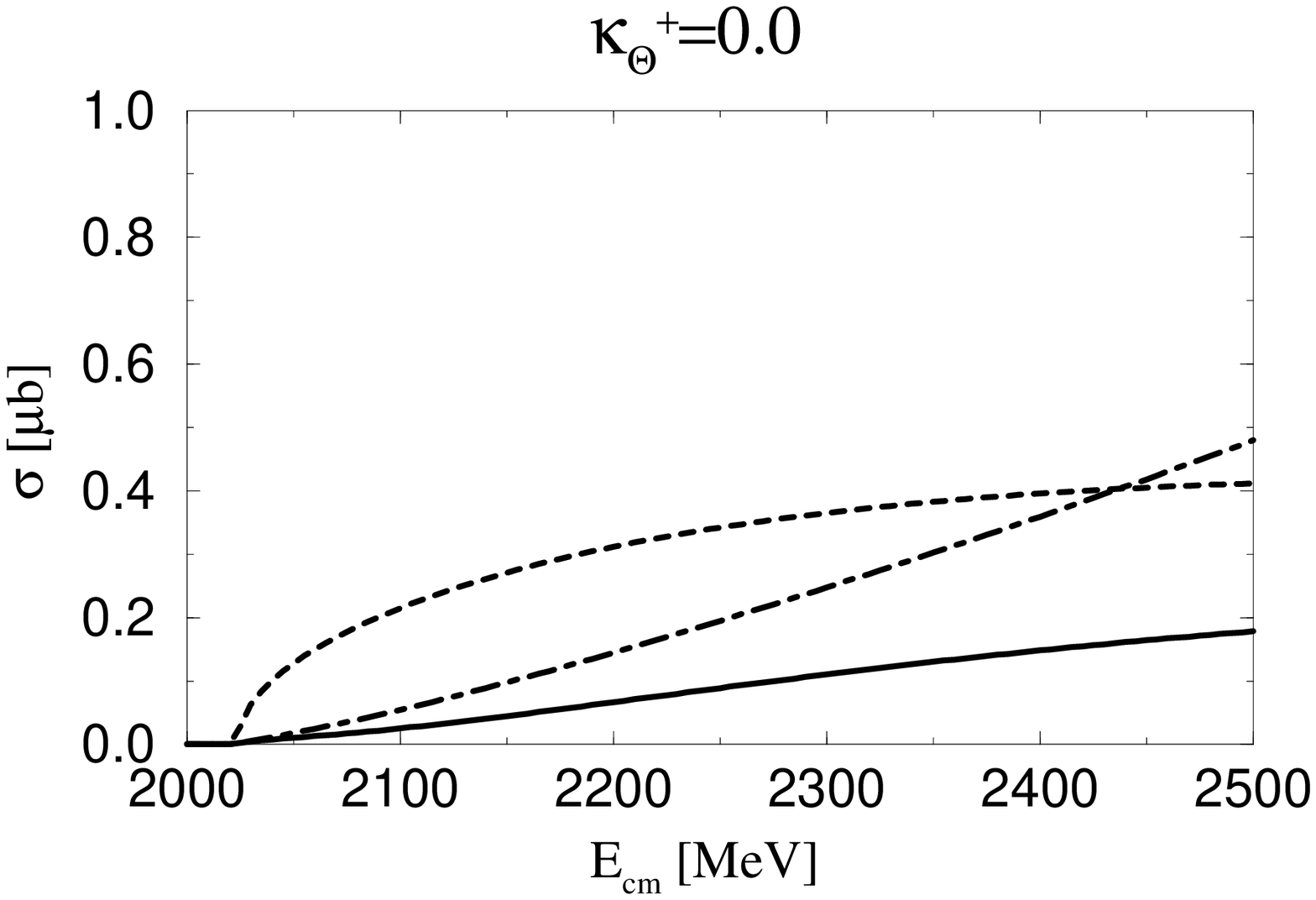}}
\resizebox{5.5cm}{4cm}{\includegraphics{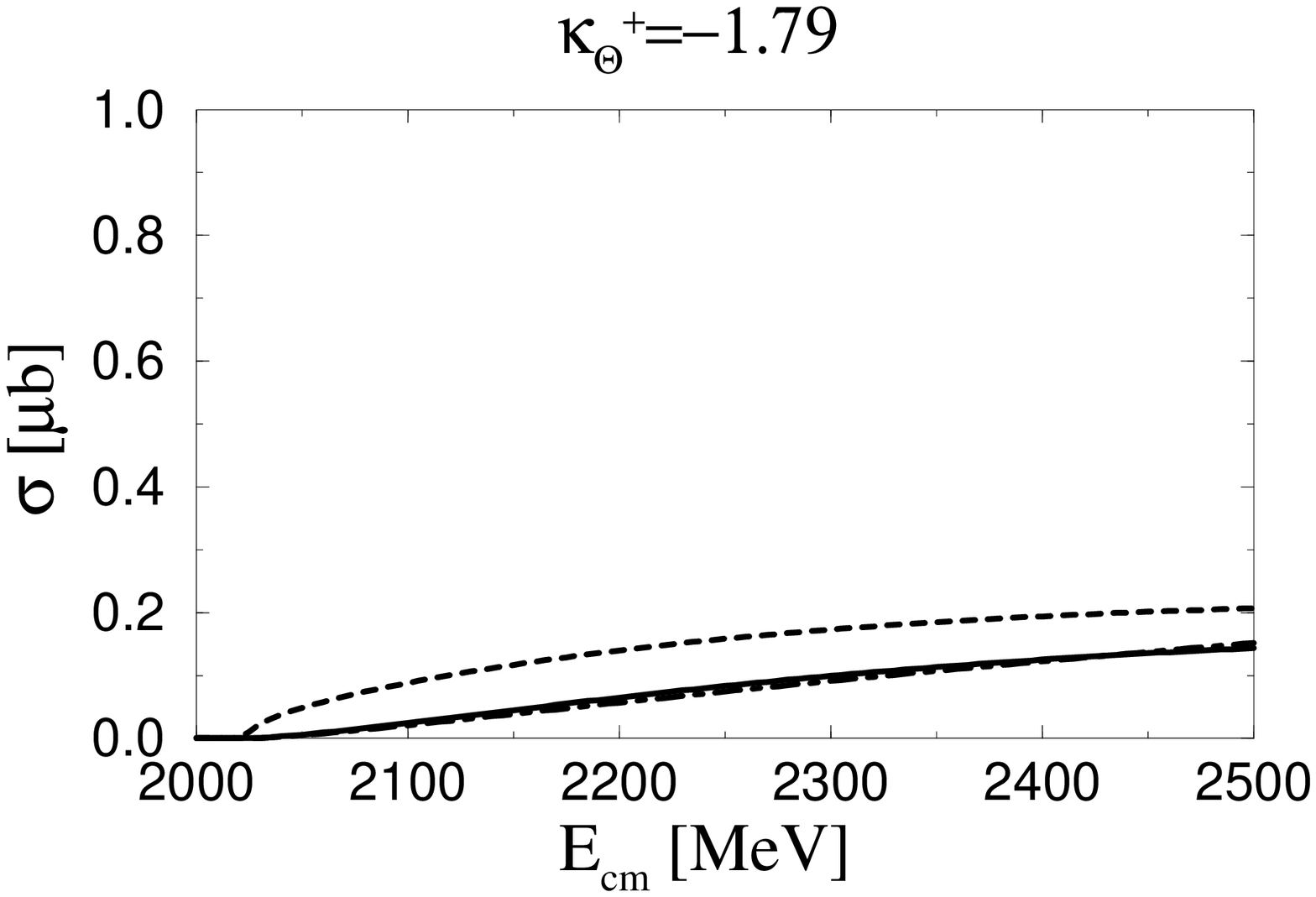}}
\end{tabular}
\caption{Total cross sections for three different $\kappa_{\Theta}$'s for
$J^P=1/2^{-}$. Solid, dotted and dot-dashed lines are for HM, 
PS and PV, respectively.}
\label{TCSm}
\end{figure}

\begin{figure}[ht]
\begin{tabular}{ccc}
\resizebox{5.5cm}{4cm}{\includegraphics{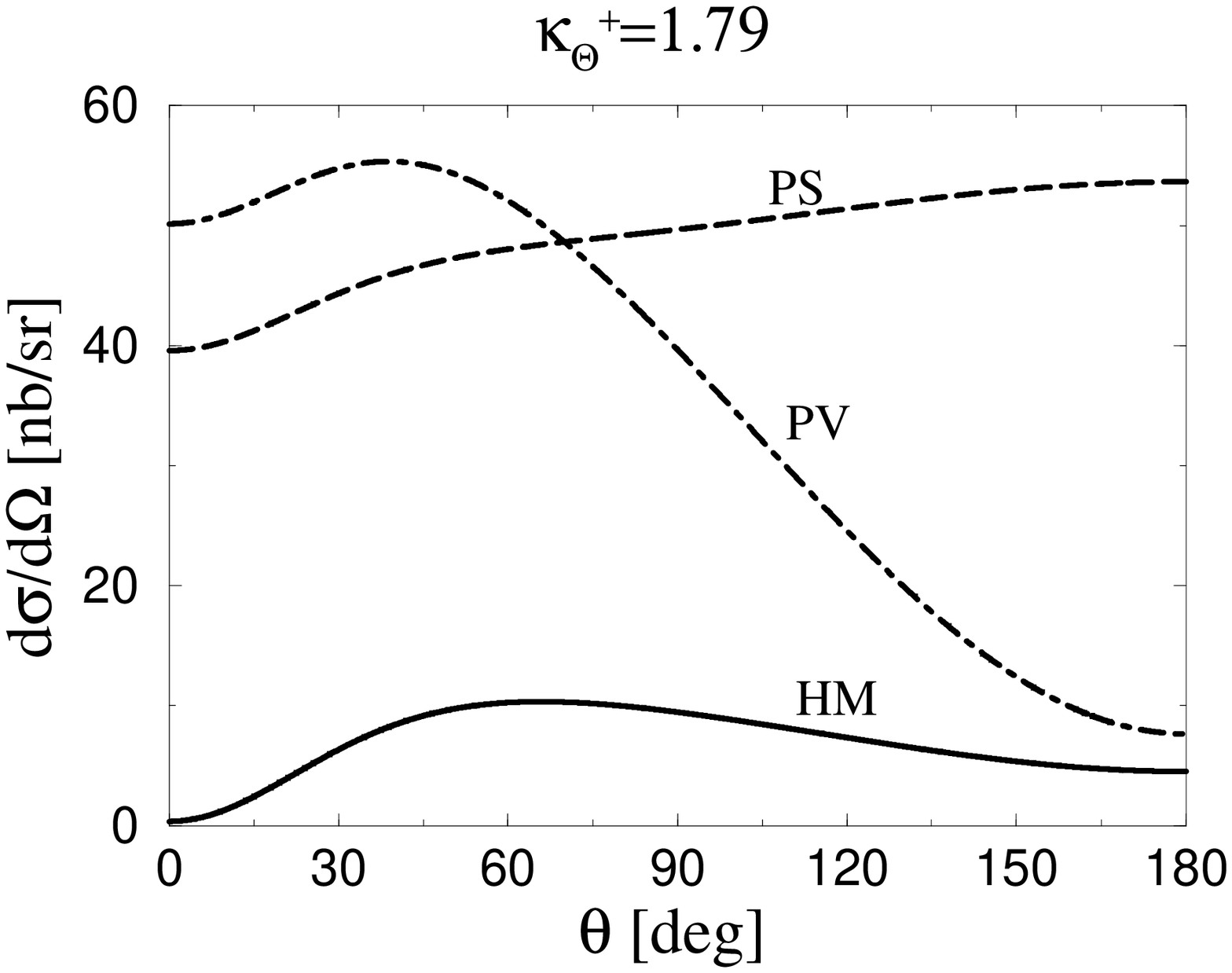}}
\resizebox{5.5cm}{4cm}{\includegraphics{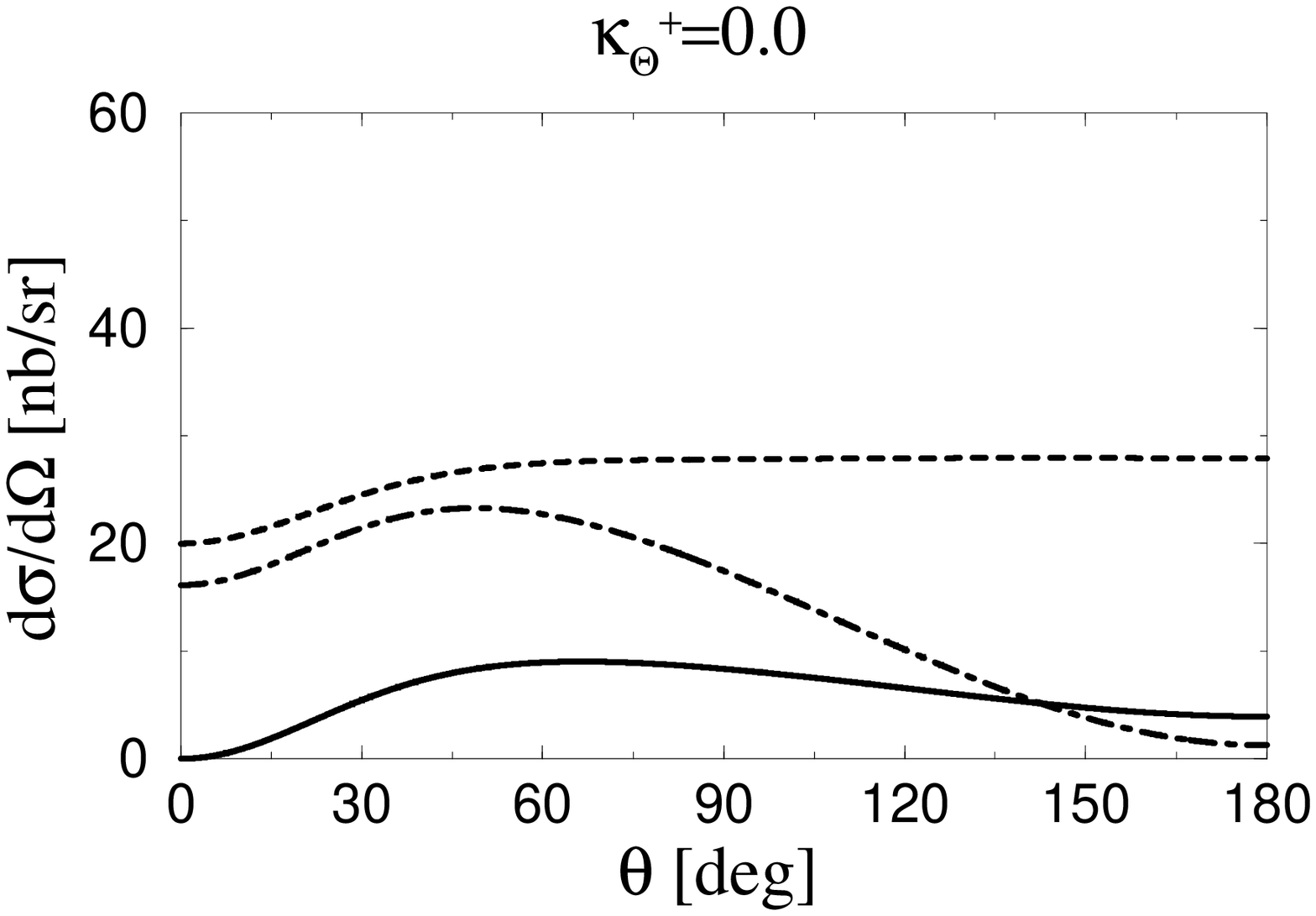}}
\resizebox{5.5cm}{4cm}{\includegraphics{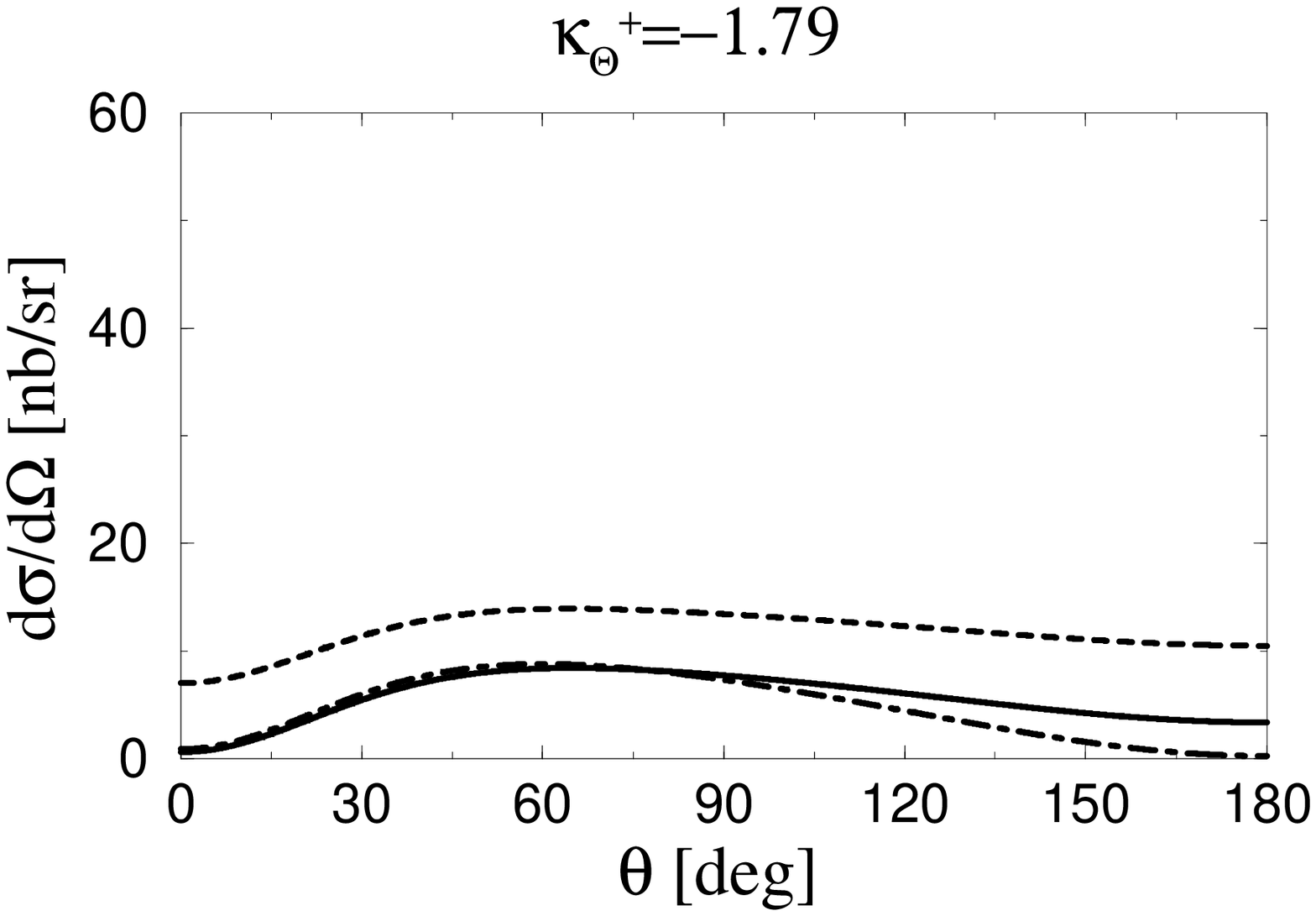}}
\end{tabular}
\caption{Differential cross sections  for three different
$\kappa_{\Theta}$'s for $J^P=1/2^{-}$ at ${\rm E_{cm}}$=2.25
GeV. Solid, dotted and 
dot-dashed lines are for HM, 
PS and PV, respectively.}
\label{DCSm}
\end{figure}

Next, in Figs.~\ref{TCSm} and \ref{DCSm} we show the results for a
negative parity $\Theta^+$.   
The total cross sections are much smaller than 
the case of the positive parity by about a factor $1/10 \sim 1/5$. 
This can be understood as follows.  
For instance, as we have seen in the PS scheme the Born diagrams 
of s and u-channel processes are dominant.  
In this case, the $KN\Theta$ coupling for positive parity 
$\Theta^+$ is $\gamma_5$ (for negative parity it is 1 as in 
the non-relativistic case).  
The Born diagram contains also the electromagnetic 
coupling proportional to $\vec \epsilon \cdot \vec \gamma$, 
where $\vec \epsilon$ is a spatial part of the photon polarization 
vector.    
From these structures, one can see that the amplitude of the 
direct Born term is of order 1 for the positive parity $\Theta^+$, 
while 
it is of order $p/M$ for the negative parity $\Theta^+$.  
The propagation of the negative energy neutron 
in the Z-graph type contributes to the 
term of order 1 for the positive parity $\Theta^+$
while there is no such a leading order term for the negative 
parity $\Theta^+$.  
Similar suppression is verified for the PV scheme in the 
KR term as it should be.  
We can also verify that t-channel diagram is also 
suppressed for a negative parity $\Theta^+$.  
In the non-relativistic limit and 
for on-shell particles, 
the $KN\Theta$ vertex has a structure of $\vec{\sigma} \cdot \vec{q}$ 
for the positive parity 
$\Theta^+$, while 1 for the negative parity $\Theta^+$, 
where $\vec q$ is an exchanged momentum.  
The strength of the coupling is fixed from the 
decay of $\Theta^+$ in which the momentum is 
$q \sim 256$ MeV.  
In the t-channel diagram, however, the 
relevant momentum is about 1 GeV, which suppresses 
the amplitude for the negative parity 
$\Theta^+$ by the factor $1/4$.  

Once again, the total cross section depends on the 
lagrangian scheme and on $\kappa_{\Theta}$.  
The largest total cross section for the negative parity 
$\Theta^+$ is obtained in the PV scheme 
for $\kappa_{\Theta} = + 1.79$.    
But if we take the result of the HM and including 
the suppression factor of 18 \% for a 
reasonable estimate, we obtain:  
20 nb (for $\kappa_{\Theta} = + 1.79)$, 
16 [nb] (for $\kappa_{\Theta} = 0)$, 
14 [nb] (for $\kappa_{\Theta} = - 1.79)$.

We have also shown in Fig.~\ref{DCSm} 
differential cross sections for the case of a negative parity $\Theta^+$.  
In this case, it is interesting to see that the cross sections are 
suppressed at the forward angle in all cases 
(more precisely, the results for PS with 
$\kappa_{\Theta} = 1.79$ and 0 show only a local 
minimum at the forward angle.).  
This is a sharp contrast to the result of the 
positive parity $\Theta^+$.

Finally, we comment on the possible range of the 
decay width of $\Theta^+$.  
In DIANA collaboration~\cite{Barmin:2003es},
$\Gamma(\Theta \rightarrow KN)$  was measured to be $<$ 9 MeV. 
This result is consistent with the remark 
in Ref.~\cite{Diakonov:1997mm} that $\Gamma$
could be smaller than 15 MeV. 
In fact, by employing
$g^{(0)}_{A}\sim0.37$~\cite{Blotz:wi}, $\Gamma$ becomes about 10 MeV.  
If we use a smaller coupling constant corresponding to this 
narrower width, cross sections are reduced.

\section{Summary and discussions}

In table~\ref{summ}, we have summarized our main conclusion of the 
present study, where 
we present the mean total cross sections estimated in the HM and with
a suppression  
factor of 18 \% expected from a form factor.  
The $KN\Theta$ coupling constants are determined 
by $\Gamma_\Theta \sim 15$ MeV.  
Although we have started from the PS and PV schemes of the 
effective lagrangians (this was convenient to make a consistency 
check of our numerical calculations), in order to obtain a better 
energy dependence of the 
total cross section as expected from the $K\Lambda$ production, we 
have shown the results of the HM.  
For the magnetic moments of $\Theta^+$, a prediction in the chiral
soliton model and crude estimation using other models scatter 
around  $-0.7 \lsim \kappa_{\Theta} \lsim 0.3$ in units of $\Theta$
magneton. When a value of $\Theta$ not far from zero, we may use the
two features to determine the parity of  
$\Theta^+$; 
one is the total cross sections which differ by about a factor 
ten for the positive and negative parity cases, 
and the other is angular dependence which shows a forward 
peak structure for the positive parity $\Theta^+$ but with 
a suppressed one for the negative parity $\Theta^+$ especially 
near the threshold region.

\begin{table}[tbh]
\centering
                \label{summ}
                \begin{minipage}{12cm}
                \caption{Summary of various cross sections. Numbers
                for total cross sections are averaged in the energy
                range $E_{CM} \lsim 2.5$ GeV. In the third and fifth
                column, FP denotes ``forward peak".   
                See the text for more details.}.  
                \end{minipage}
        \vspace*{0.3cm}
\begin{tabular}{r |  c  c c c c c  c}
        \hline
        & \multicolumn{3}{c}{Positive parity}  &
        & \multicolumn{3}{c}{Negative parity} 
        \\
                 & Total $\sigma$ & & Angular dep. & 
                 & Total $\sigma$ & & Angular dep. \\
        \hline
        $\kappa_\Theta = +1.79$ 
        & 240 nb & & Weak & & 20 nb & & FP suppressed
        \\
        0
        & 140 nb & & FP enhanced & & 16 nb &  & "
        \\
        $ -1.79$
        & 100 nb & & FP enhanced & & 14 nb & & "
        \\
        \hline
    \end{tabular}
\end{table}

Both observables need more careful analysis when compared 
with data, since they depend very much on  
the lagrangian schemes and the unknown parameters.  
In particular, the total cross sections are 
very sensitive to the form factor as it should be.  
On the other hand, the qualitative structure of the 
angular dependence seems rather stable as long as the HM scheme is adopted.  

As reported by the DIANA group, if the decay width of 
$\Theta^+$ would smaller than 9 MeV, say for instance 
5 MeV, the total cross section is reduced by the factor 
three and the estimated $\sigma(\gamma n \to K^-\Theta^+)$
would be of order 10 nb.  
If we assume that the parity of $\Theta^+$ is negative, the 
total cross section would be even smaller which would be 
of order one nb.  
Definitely, we need to study more both theoretically and experimentally
in order to understand the nature of the pentaquark state $\Theta^{+}$.

\section*{Acknowledgments}
 We thank Takahashi Nakano for stimulating
discussions. We also thank Hiroshi Toki for useful comments.
The work of HCK is supported by the Korean Research Foundation
(KRF--2000--015--DP0069).  The works of SINam has been supported by
scholarship of the Ministry of Education, Sciences, 
Sports and Culture of Japan.

\end{document}